# SEMI-LEPTONIC FORM-FACTORS FROM LATTICE QCD[*]

Tanmoy Bhattacharya and Rajan Gupta
*T-8 Group, MS B285, Los Alamos National Laboratory, Los Alamos,
New Mexico 87545, U. S. A.*

ABSTRACT

We present results for semi-leptonic form-factors obtained on a statistical sample of 66 $32^3 \times 64$ lattices at $\beta = 6.0$ using quenched Wilson fermions. We find $f_+^{D \to K l\nu}(q^2 = 0) = 0.73 \pm 0.06$, $A_2/A_1(D \to K^* l\nu) = 0.72 \pm 0.22$, $V/A_1(D_s \to \phi l\nu) = 1.91 \pm 0.04$, and $A_2/A_1(D_s \to \phi l\nu) = 0.68 \pm 0.09$, where the error estimate includes statistical errors and errors due to extrapolation to $q^2 = 0$ and to physical values of $(m_u + m_d)/2$ and $m_s$. The remaining sources of systematic errors are those due to $O(a)$ discretization errors and those due to quenching, which our results indicate may be small. We also comment on the validity of pole-dominance in these form-factors.

## 1. INTRODUCTION

Exclusive semi-leptonic decays of $D$ and $B$ mesons provide the cleanest measurements of the CKM quark mixing matrix. For example, the decay rate for $D \to K l\nu$,

$$\frac{d\Gamma^{D \to Kl\nu}}{dq^2} = \frac{G_F^2}{24\pi^3} |V_{cs}|^2 p_K^3 f_+^2(q^2), \quad (1)$$

depends on kinematic factors, a single CKM matrix element $V_{cs}$, and the form-factor $f_+(q^2)$. To extract CKM matrix elements from such processes requires non-perturbative calculations of the form-factors as they encapsulate all strong interaction effects. In this talk we report on results obtained from numerical simulations of lattice QCD.

## 2. LATTICE PARAMETERS

The results presented here have been obtained using the following lattice parameters. The $32^3 \times 64$ gauge lattices were generated at $\beta = 6.0$ using the combination 5 over-relaxed (OR) sweeps followed by 1 Metropolis sweep. Quark propagators are calculated on lattices separated by 2000 OR sweeps using the simple Wilson action. Periodic boundary conditions are used in all 4 directions, both during lattice update and propagator calculation. Quark propagators have been calculated using one version of Wuppertal smeared sources at $\kappa = 0.135$ ($C$), $0.153$ ($S$), $0.155$ ($U_1$), $0.1558$ ($U_2$), and $0.1563$ ($U_3$). These quark masses correspond to pseudoscalar mesons of mass 2800, 980, 700, 550 and 440 $MeV$ respectively using $1/a = 2.25(10) GeV$ set by $m_\rho$. On each of the 66 configurations we make two independent measurements of the form-factors, which we average before doing the statistical analysis using the jackknife method.

---

[*] Talk presented by R. Gupta at DPF94.

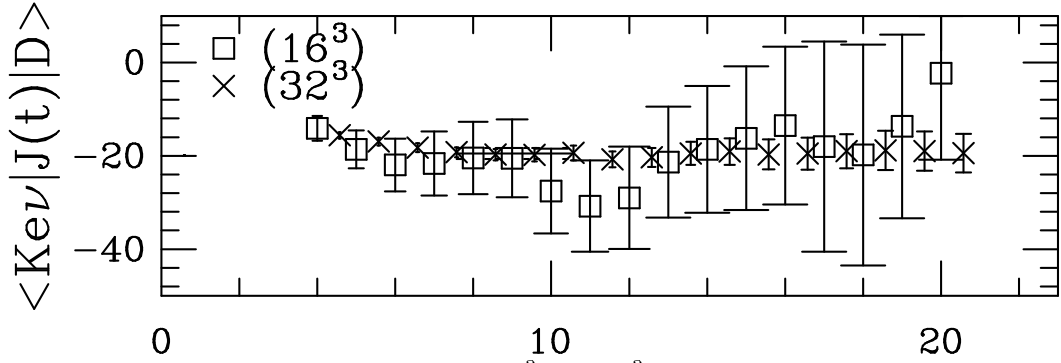
Fig. 1. Comparison of signal on $16^3$ and $32^3$ lattices as a function of $t$.

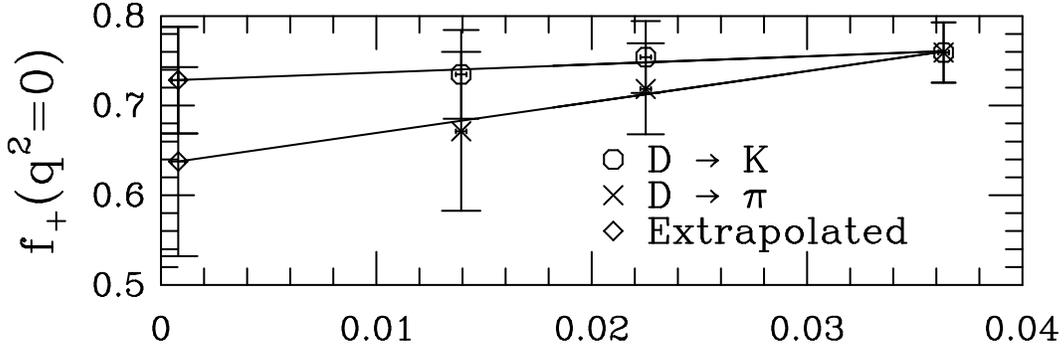
Fig. 2. Extrapolation of $f_+^K$ and $f_+^\pi$ in the light quark mass $\bar{m}a$.

Our procedure for extracting form-factors is very similar to that proposed by Lubicz el al.,[1] and a detailed paper is under preparation. The $D$ meson is created at $\vec{p} = (0,0,0)$ and the momentum inserted by the current is carried by the final kaon. The five values of momenta analyzed are $\vec{p} = (0,0,0)$, $\vec{p} = (1,0,0)$, $\vec{p} = (1,1,0)$, $\vec{p} = (1,1,1)$, and $\vec{p} = (2,0,0)$ in units of $\pi/16a$. These correspond to roughly 0, 440, 625, 765, and 880 MeV respectively.

The use of large lattices to study form-factors leads to a dramatic improvement in reliability. In Fig. 1 we show a comparison of the signal in $\langle K \mid V_i \mid D \rangle$ with $\vec{p} = (\pi/8a, 0, 0)$ for our current data set (132 measurements) with a previous study using 35 $16^3$ lattices. The reduction in errors by a factor of $\approx 5$ is consistent with the increase in statistics and lattice volume. In addition, the larger lattice allows measurements at three smaller values of non-zero momentum transfer, for which the signal is even better. These points bracket $q^2 = 0$ and allow a reliable extraction of $f(q^2 = 0)$, which we do in two ways. Our best fit uses a two parameter fit to the pole-dominance ansatz $f(q^2) = f(0)/(1 - q^2/\mathcal{M}^2)$. In the second method we fix the pole mass $\mathcal{M}$ to its lattice measured value. The relative merits of the two methods are discussed below.

We take $\kappa = 0.135$ as the physical charm quark. The ratio $m_\pi^2/m_K^2$ fixes the strange quark at $\kappa = 0.1550(2)$. The three light quarks $U_1 - U_3$ are used to extrapolate the $q^2 = 0$ data to the physical value of $\bar{m} \equiv (m_u + m_d)/2$ (fixed by the experimental ratio $m_\pi^2/m_\rho^2$) assuming that the form-factors depend linearly on the light quark mass. For example, the extrapolation of $f_+$ is shown in Fig. 2. Also, to calculate $A_2/A_1$, $V/A_1$ etc, the ratios of form-factors are taken at the very beginning of the jackknife process.

|  | $f_+(q^2=0)$ | | | $f_0(q^2=0)$ | |
| --- | --- | --- | --- | --- | --- |
|  | (a) | (b) | EXPT. | (a) | (b) |
| $D \to Kl\nu$ | 0.72(5) | 0.81(3) | 0.77(4) | 0.73(4) | 0.73(2) |
| $D \to \pi l\nu$ | 0.64(9) | 0.75(4) |  | 0.63(6) | 0.65(3) |
| $(D \to \pi l\nu)/(D \to Kl\nu)$ | 0.88(6) | 0.93(2) | $1.29 \pm 0.21 \pm 0.11$ | 0.88(4) | 0.89(2) |

Table 1. form-factors, $f_+$ and $f_0$, extracted using (a) best fit and (b) lattice pole masses.

|  | $V$ | | $A_1$ | | $A_2$ | |
| --- | --- | --- | --- | --- | --- | --- |
|  |  | exp. |  | exp. |  | exp. |
| $D \to K^* l\nu$ | 1.24( 8) | 1.16(16) | 0.66(3) | 0.61(5) | 0.45(19) | 0.45(9) |
| $D \to \rho l\nu$ | 1.08(12) |  | 0.56(4) |  | 0.19(24) |  |
| $D_s \to \phi l\nu$ | 1.29( 5) |  | 0.66(1) |  | 0.46( 8) |  |

Table 2. Estimates for vector form-factors using fits with lattice pole masses.

## 3. FINAL RESULTS

The results for the decay $D \to Kl\nu$ are given in Table 1. Present errors preclude a serious test of the pole-dominance hypothesis even though the best fit value for $\mathcal{M}_{1^-}$ is about $10-20\%$ below the mass measured on the lattice and $20-30\%$ below the known experimental values. Since $f_+$ is known from experiments,[2] one can regard the lattice measurements as providing a measure of systematic errors due to quenching and lattice discretization that we cannot otherwise estimate. The data for $f_+$ in Table 1 suggest that these are small, *i.e.* at the 10% level. Our best estimate for $(D \to \pi l\nu)/(D \to Kl\nu) = 0.88(6)$ lies at the lower end of the range of experimental values.[2]

In the case of the vector final states we find that the estimates for $A_2$ are not very stable for $\vec{p} = (1,1,1)$ and $(2,0,0)$, making a two free parameter fit (best fit method) unreliable. However, the point $\vec{p} = (1,1,0)$ lies very close to the desired limit $q^2 = 0$, and can be used as a consistency check. With this criterion we find that the pole fits give reasonable estimates. For $V$ and $A_1$ the two kinds of fits give consistent estimates, therefore in Table 2 we give pole fit results as our best estimates for all three form-factors. The results for $D \to K^* l\nu$ are in surprisingly good agreement with the averaged experimental values.[2] The form-factors for $D \to \rho l\nu$ are consistently smaller and we find little difference, qualitatively or quantitatively, between the two final states $K^*$ and $\phi$. The experimental errors in $D_s \to \phi l\nu$ are too large (see summary talk by Janis McKenna in these proceedings) to make a meaningful comparison.

We gratefully acknowledge the tremendous support provided by the ACL at Los Alamos, and NCSA at Urbana-Champaign. These calculations have been done on CM5 parallel supercomputers as part of the DOE HPCC Grand Challenges program (at ACL) and an NSF Metacenter allocation (at NCSA).